\documentstyle[12pt]{article}
\def\be {\begin{equation}}
\def\ee {\end{equation}}
\def\ba {\begin{eqnarray}}
\def\ea {\end{eqnarray}}
\newcommand{\bq}{\begin{eqnarray}}
\newcommand{\eq}{\end{eqnarray}}

\def\bi {\begin{itemize}}
\def\ei {\end{itemize}}
\begin{document}
\def\bea{\begin{eqnarray}}
\def\eea{\end{eqnarray}}
\title{\bf {Interacting holographic generalized Chaplygin gas model }}
 \author{M.R. Setare  \footnote{E-mail: rezakord@ipm.ir}
  \\ {Department of Science,  Payame Noor University. Bijar, Iran}}
\date{\small{}}
\maketitle
\begin{abstract}
In this paper we consider a correspondence between the holographic
dark energy density and interacting generalized  Chaplygin gas
energy density in FRW universe.  Then we reconstruct the potential
of the
 scalar field which describe the generalized  Chaplygin cosmology.
 \end{abstract}

\newpage
\section{Introduction}
The accelerated expansion that based on recent astrophysical data
\cite{exp}, our universe is experiencing  is today's most important
problem of cosmology. Missing energy density - with negative
pressure - responsible for this expansion has been dubbed Dark
Energy (DE). Wide range of scenarios have been proposed to explain
this acceleration while most of them can not explain all the
features of universe or they have so many parameters that makes them
difficult to fit. The models which have been discussed widely in
literature are those which consider vacuum energy (cosmological
constant) \cite{cosmo} as DE, introduce fifth elements and dub it
quintessence \cite{quint} or scenarios named phantom \cite{phant}
with $w<-1$ , where $w$ is parameter of state.

An approach to the problem of DE arises from holographic principle
that states that the number of degrees of freedom related directly
to entropy scales with the enclosing area of the system. It was
shown by 'tHooft and Susskind \cite{hologram} that effective local
quantum field theories greatly overcount degrees of freedom because
the entropy scales extensively for an effective quantum field theory
in a box of size $L$ with UV cut-off $ \Lambda$. As pointed out by
\cite{myung}, attempting to solve this problem, Cohen {\it et al.}
showed \cite{cohen} that in quantum field theory, short distance
cut-off $\Lambda$ is related to long distance cut-off $L$ due to the
limit set by forming a black hole. In other words the total energy
of the system with size $L$ should not exceed the mass of the same
size black hole i.e. $L^3 \rho_{\Lambda}\leq LM_p^2$ where
$\rho_{\Lambda}$ is the quantum zero-point energy density caused by
UV cutoff $\Lambda$ and $M_P$ denotes Planck mass ( $M_p^2=1/{G})$.
The largest $L$ is required to saturate this inequality. Then its
holographic energy density is given by $\rho_{\Lambda}= 3c^2M_p^2/
L^2$ in which $c$ is free dimensionless parameter and coefficient 3
is for convenience. Based on cosmological state of holographic
principle, proposed by Fischler and Susskind \cite{fischler}, the
Holographic model of Dark Energy (HDE) has been proposed and studied
widely in the
 literature \cite{miao,HDE}.\\
Some experimental data has implied that our universe is not a
perfectly flat universe and recent papers have favored the universe
with spatial curvature \cite{{wmap},{ws}}. As a matter of fact, we
want to remark that although it is believed that our universe is
flat, a contribution to the Friedmann equation from spatial
curvature is still possible if the number of e-foldings is not very
large \cite{miao2}. Defining the appropriate distance, for the case
of non-flat universe has another story. Some aspects of the problem
has been discussed in \cite{miao2,guberina}. In this case, the event
horizon can not be considered as the system's IR cut-off, because
for instance, when the dark energy is dominated and $c=1$, where $c$
is a positive constant, $\Omega_\Lambda=1+ \Omega_k$, we find $\dot
R_h<0$, while we know that in this situation we must be in de Sitter
space with constant EoS. To solve this problem, another distance is
considered- radial size of the event horizon measured on the sphere
of the horizon, denoted by $L$- and the evolution of holographic
model of dark energy in non-flat universe is investigated.
\\
It is fair to claim that simplicity and reasonability of HDE
provides
 more reliable frame to investigate the problem of DE rather than other models
proposed in the literature\cite{cosmo,quint,phant}. For instance the
coincidence or "why now" problem is easily solved in some models of
HDE based on this fundamental assumption that matter and holographic
dark energy do not conserve separately, but the matter energy
density decays into the holographic energy density \cite{intde}. In
fact a suitable evolution of the Universe is obtained when, in
addition to the holographic dark energy, an interaction (decay of
dark energy to matter) is assumed.\\
 In a
very interesting paper Kamenshchik, Moschella, and Pasquier
\cite{kmp}have studied a homogeneous model based on a single fluid
obeying the Chaplygin gas equation of state \be \label{chp}
P=\frac{-A}{\rho} \ee where $P$ and $\rho$ are respectively pressure
and energy density in comoving reference frame, with $\rho> 0$; $A$
is a positive constant. This equation of state has raised  a certain
interest \cite{jac} because of its many interesting and, in some
sense, intriguingly unique features. Some possible motivations for
this model from the field theory points of view are investigated in
\cite{a}. The Chaplygin gas emerges as an effective fluid associated
with d-branes \cite{b} and can also be obtained from the Born-Infeld
action \cite{c}.\\
Inserting the equation of state (\ref{chp}) into the relativistic
energy conservation equation, leads to a density evolving as \be
\label{enerd}\rho_{\Lambda}=\sqrt{A+\frac{B}{a^{6}}} \ee where $B$
is an integration constant.\\
In present paper, using the generalized Chaplygin gas model of dark
energy, we obtain equation of state for interacting Chaplygin gas
energy density in non-flat universe. The current available
observational data imply that the dark energy behaves as
phantom-type dark energy, i.e. the equation-of-state of dark energy
crosses the cosmological-constant boundary $w=-1$ during the
evolution history. We show this phantomic description of the
interacting generalized Chaplygin gas dark energy in non-flat
universe, and reconstruct the potential of the phantom scalar field.
Finally we introduce the squared speeds of sound for generalized
Chaplygin gas and interacting holographic fluid. We find that the
squared speed for generalized Chaplygin gas is negative when,
$\alpha<0$, $A> 0$ or $\alpha >0$, $A <0$. Also we show that the
squared speed for interacting holographic fluid is negative when
choosing $c = 0.84$ , and taking $ \Omega_{\Lambda}= 0.73$, $
\Omega_{k} = 0.01$ for the present time.
\section{ Interacting generalized Chaplygin gas model}
In this section we obtain the equation of state for the generalized
Chaplygin gas when there is an interaction between generalized
Chaplygin gas energy density $\rho_{\Lambda}$ and a Cold Dark
Matter(CDM) with $w_{m}=0$. The continuity equations for dark energy
and CDM are
\begin{eqnarray}
\label{2eq1}&& \dot{\rho}_{\rm \Lambda}+3H(1+w_{\rm \Lambda})\rho_{\rm \Lambda} =-Q, \\
\label{2eq2}&& \dot{\rho}_{\rm m}+3H\rho_{\rm m}=Q.
\end{eqnarray}
The interaction is given by the quantity $Q=\Gamma \rho_{\Lambda}$.
This is a decaying of the generalized Chaplygin gas component into
CDM with the decay rate $\Gamma$. Taking a ratio of two energy
densities as $r=\rho_{\rm m}/\rho_{\rm \Lambda}$, the above
equations lead to
\begin{equation}
\label{2eq3} \dot{r}=3Hr\Big[w_{\rm \Lambda}+
\frac{1+r}{r}\frac{\Gamma}{3H}\Big]
\end{equation}
 Following Ref.\cite{Kim:2005at},
if we define
\begin{eqnarray}\label{eff}
w_\Lambda ^{\rm eff}=w_\Lambda+{{\Gamma}\over {3H}}\;, \qquad w_m
^{\rm eff}=-{1\over r}{{\Gamma}\over {3H}}\;.
\end{eqnarray}
Then, the continuity equations can be written in their standard form
\begin{equation}
\dot{\rho}_\Lambda + 3H(1+w_\Lambda^{\rm eff})\rho_\Lambda =
0\;,\label{definew1}
\end{equation}
\begin{equation}
\dot{\rho}_m + 3H(1+w_m^{\rm eff})\rho_m = 0\; \label{definew2}
\end{equation}
We consider the non-flat Friedmann-Robertson-Walker universe with
line element
 \be\label{metr}
ds^{2}=-dt^{2}+a^{2}(t)(\frac{dr^2}{1-kr^2}+r^2d\Omega^{2}).
 \ee
where $k$ denotes the curvature of space k=0,1,-1 for flat, closed
and open universe respectively. A closed universe with a small
positive curvature ($\Omega_k\sim 0.01$) is compatible with
observations \cite{ {wmap}, {ws}}. We use the Friedmann equation to
relate the curvature of the universe to the energy density. The
first Friedmann equation is given by
\begin{equation}
\label{2eq7} H^2+\frac{k}{a^2}=\frac{1}{3M^2_p}\Big[
 \rho_{\rm \Lambda}+\rho_{\rm m}\Big].
\end{equation}
Define as usual
\begin{equation} \label{2eq9} \Omega_{\rm
m}=\frac{\rho_{m}}{\rho_{cr}}=\frac{ \rho_{\rm
m}}{3M_p^2H^2},\hspace{1cm}\Omega_{\rm
\Lambda}=\frac{\rho_{\Lambda}}{\rho_{cr}}=\frac{ \rho_{\rm
\Lambda}}{3M^2_pH^2},\hspace{1cm}\Omega_{k}=\frac{k}{a^2H^2}
\end{equation}
Now we can rewrite the first Friedmann equation as
\begin{equation} \label{2eq10} \Omega_{\rm m}+\Omega_{\rm
\Lambda}=1+\Omega_{k}.
\end{equation}
Using Eqs.(\ref{2eq9},\ref{2eq10}) we obtain following relation for
ratio of energy densities $r$ as
\begin{equation}\label{ratio}
r=\frac{1+\Omega_{k}-\Omega_{\Lambda}}{\Omega_{\Lambda}}
\end{equation}
In the generalized Chaplygin gas approach \cite{c}, the equation of
state to (\ref{chp}) is generalized to \be \label{chpge}
P_{\Lambda}=\frac{-A}{\rho_{\Lambda}^{\alpha}} \ee The above
equation of state leads to a density evolution as \be \label{denge}
\rho_{\Lambda}=[A+\frac{B}{a^{3(1+\alpha)}}]^{\frac{1}{1+\alpha}}
\ee By considering the above equations, one can find
\begin{equation}\label{stateq}
w_{\rm
\Lambda}=\frac{P_{\Lambda}}{\rho_{\Lambda}}=\frac{-A}{a^{3(1+\alpha)}[A+B
a^{-3(1+\alpha)}]}.
\end{equation}
From Eqs.(\ref{eff}, \ref{stateq}), we have the effective equation
of state as
\begin{equation} \label{3eq401}
w_{\rm \Lambda}^{eff}=\frac{-A}{a^{3(1+\alpha)}[A+B
a^{-3(1+\alpha)}]}+\frac{\Gamma}{3H}.
\end{equation}
Here as in Ref.\cite{WGA}, we choose the following relation for
decay rate
\begin{equation}\label{decayeq}
\Gamma=3b^2(1+r)H
\end{equation}
with  the coupling constant $b^2$. Now using the  definition of
generalized Chaplygin gas energy density $\rho_{\rm \Lambda}$, and
using $\Omega_{\Lambda}$, we can rewrite Eq.(\ref{3eq401}) as
\begin{equation}\label{stateq2}
w_{\rm \Lambda}^{eff}=\frac{-A}{(3M_{p}^{2} H^2 \Omega_{\rm
\Lambda})^{1+\alpha}}+\frac{b^2(1+\Omega_{k})}{\Omega_{\rm \Lambda}}
\end{equation}
Now we suggest a correspondence between the holographic dark energy
scenario and the generalized Chaplygin gas dark energy model.\\
In non-flat universe, our choice for holographic dark energy density
is
 \be \label{holoda}
  \rho_\Lambda=3c^2M_{p}^{2}L^{-2}.
 \ee
$L$ is defined as the following form\cite{miao2}:
\begin{equation}\label{leq}
 L=ar(t),
\end{equation}
here, $a$, is scale factor and $r(t)$ is relevant to the future
event horizon of the universe. Given the fact that
\begin{eqnarray}
\int_0^{r_1}{dr\over \sqrt{1-kr^2}}&=&\frac{1}{\sqrt{|k|}}{\rm
sinn}^{-1}(\sqrt{|k|}\,r_1)\nonumber\\
&=&\left\{\begin{array}{ll}
\sin^{-1}(\sqrt{|k|}\,r_1)/\sqrt{|k|},\ \ \ \ \ \ &k=1,\\
r_1,&k=0,\\
\sinh^{-1}(\sqrt{|k|}\,r_1)/\sqrt{|k|},&k=-1,
\end{array}\right.
\end{eqnarray}
one can easily derive \be \label{leh} L=\frac{a(t) {\rm
sinn}[\sqrt{|k|}\,R_{h}(t)/a(t)]}{\sqrt{|k|}},\ee where $R_h$ is the
future event horizon given by \be
  R_h= a\int_t^\infty \frac{dt}{a}=a\int_a^\infty\frac{da}{Ha^2}
 \ee
 By considering  the definition of
holographic energy density $\rho_{\rm \Lambda}$, one can find
\cite{{set1},{set2}}:
\begin{equation}\label{stateq4}
w_{\rm \Lambda}=-[\frac{1}{3}+\frac{2\sqrt{\Omega_{\rm
\Lambda}}}{3c}\frac{1}{\sqrt{|k|}}\rm
cosn(\sqrt{|k|}\,R_{h}/a)+\frac{\Gamma}{3H}].
\end{equation}
 where
\begin{equation}
\frac{1}{\sqrt{|k|}}{\rm cosn}(\sqrt{|k|}x)
=\left\{\begin{array}{ll}
\cos(x),\ \ \ \ \ \ &k=1,\\
1,&k=0,\\
\cosh(x),&k=-1.
\end{array}\right.
\end{equation}
Substituting Eq.(\ref{decayeq}) into Eq.(\ref{stateq}), and using
Eq.(\ref{eff}) one can find
\begin{equation} \label{3eq402}
w_{\rm \Lambda}^{eff}=-\frac{1}{3}-\frac{2\sqrt{\Omega_{\rm
\Lambda}-c^2\Omega_{k}}}{3c}.
\end{equation}
If we establish the correspondence between the holographic dark
energy and generalized Chaplygin gas energy density, then using
Eqs.(\ref{denge}, \ref{holoda})we have \be
\label{aco}A=(3c^2M_{p}^{2}L^{-2})^{1+\alpha}-\frac{B}{a^{3(1+\alpha)}}
\ee Using definitions
$\Omega_{\Lambda}=\frac{\rho_{\Lambda}}{\rho_{cr}}$ and
$\rho_{cr}=3M_{p}^{2}H^2$, we get

\begin{equation}\label{hl}
HL=\frac{c}{\sqrt{\Omega_{\Lambda}}}
\end{equation}
Now, by comparing the effective equation of states (\ref{stateq2},
\ref{3eq402}) we obtain \footnote{As one can see in this case the
 $A$ and $B$ can change with time. Similar situation can arise when
 the cosmological constant has dynamic, see for example eq.(12)
 of \cite{kmp}, (see also \cite{set4}), according to this equation
 \be A=\Lambda(\Lambda+\rho_{m}) \ee therefore, if $\Lambda$ vary
 with time \cite{shap}, $A$ does not remain constant.} \be
\label{bco}A=(3M_{p}^{2} H^2 \Omega_{\rm
\Lambda})^{1+\alpha}(\frac{b^2(1+\Omega_{k})}{\Omega_{\rm
\Lambda}}+\frac{2\sqrt{\Omega_{\rm
\Lambda}-c^2\Omega_{k}}}{3c}+\frac{1}{3}) \ee Substituting the above
relation into Eq.(\ref{aco}) we have \be \label{aco1}B=(3M_{p}^{2}
H^2 \Omega_{\rm
\Lambda}a^3)^{1+\alpha}[1-(\frac{b^2(1+\Omega_{k})}{\Omega_{\rm
\Lambda}}+\frac{2\sqrt{\Omega_{\rm
\Lambda}-c^2\Omega_{k}}}{3c}+\frac{1}{3})] \ee
\section{The
correspondence between interacting generalized Chaplygin gas and
holographic phantom}
 For the non-flat universe, the authors of \cite{obsnonflat} used the data
coming from the SN and CMB to constrain the holographic dark energy
model, and got the 1 $\sigma$ fit results: $c=0.84^{+0.16}_{-0.03}$.
If we take $c=0.84$, and taking $\Omega_{\Lambda}=0.73$,
$\Omega_{k}=0.01$ for the present time, using Eq.(\ref{3eq402}) we
obtain $w_{\rm \Lambda}^{eff}=-1.007$. Also for the flat case, the
X-ray gas mass fraction of rich clusters, as a function of redshift,
has also been used to constrain the holographic dark energy model.
The main results, i.e. the 1 $\sigma$ fit values for $c$ is:
$c=0.61^{+0.45}_{-0.21}$, in this case also we obtain $w_{\rm
\Lambda}^{eff}<-1$. This implies that one can generate phantom-like
equation of state from an interacting holographic dark energy model
in flat and non-flat universe only if $c\leq 0.84$. It must be
pointed out that the choice of $c\leq 0.84$, on theoretical level,
will bring some troubles. The Gibbons-Hawking entropy will thus
decrease since the event horizon shrinks, which violates the second
law of thermodynamics as well. However, the current observational
data indicate that the parameter $c$ in the holographic model seems
smaller than 1. Now we reconstruct the phantom potential and the
dynamics of the scalar field in light of the holographic dark energy
with $c\leq 0.84$. According to the following forms of phantom
energy density and pressure \be \label{roph1}
\rho_{\Lambda}=-\frac{1}{2}\dot{\phi}^{2}+V(\phi) \ee \be
\label{roph2} P_{\Lambda}=-\frac{1}{2}\dot{\phi}^{2}-V(\phi) \ee One
can easily derive the scalar potential and kinetic energy term as
\be \label{v} V(\phi)=\frac{1}{2}(1-w_{\rm \Lambda})\rho_{\Lambda}
\ee \be \label{phi}\dot{\phi}^{2} =-(1+w_{\rm
\Lambda})\rho_{\Lambda} \ee Differenating Eq.(\ref{2eq7}) with
respect to the cosmic time $t$, one find \be
\label{hdot}\dot{H}=\frac{\dot{\rho}}{6H M_{p}^{2}}+\frac{k}{a^2}
\ee where $\rho=\rho_{m}+\rho_{\Lambda}$ is the total energy
density, now using Eqs.(\ref{2eq1}, \ref{2eq2}) \be \label{doro}
\dot{\rho}=-3H(1+w)\rho \ee  where \be
\label{weq}w=\frac{w_{\Lambda}\rho_{\Lambda}}{\rho}=\frac{\Omega_{\Lambda}w_{\Lambda}}{1+\frac{k}{a^2H^2}}
\ee Substituting $\dot{\rho}$ into Eq.(\ref{hdot}), we obtain \be
\label{weq2}
w=\frac{2/3(\frac{k}{a^2}-\dot{H})}{H^2+\frac{k}{a^2}}-1 \ee Using
Eqs.(\ref{weq}, \ref{weq2}), one can rewrite the holographic energy
equation of state as \be \label{eqes1}
w_{\Lambda}=\frac{-1}{3\Omega_{\Lambda}H^{2}}(2\dot{H}+3H^2+\frac{k}{a^2})
\ee Substituting the above $w_{\Lambda}$ into Eqs.(\ref{v},
\ref{phi}), we obtain \be \label{v1} V(\phi)=\frac{M_{p}^{2}}{2}
[2\dot{H}+3H^2(1+\Omega_{\Lambda})+\frac{k}{a^2}]\ee  \be
\label{phi2}\dot{\phi}^{2}
=M_{p}^{2}[2\dot{H}+3H^2(1-\Omega_{\Lambda})+\frac{k}{a^2}]\ee In
similar to the \cite {{odi1}, {odi2}}, we can define
$\dot{\phi}^{2}$ and $V(\phi)$ in terms of single function $f(\phi)$
as \be \label{v2} V(\phi)=\frac{M_{p}^{2}}{2}
[2f'(\phi)+3f^{2}(\phi)(1+\Omega_{\Lambda})+\frac{k}{a^2}]\ee \be
\label{phi3}1=M_{p}^{2}
[2f'(\phi)+3f^{2}(\phi)(1-\Omega_{\Lambda})+\frac{k}{a^2}]\ee In the
spatially flat case the Eqs.(\ref{v2}, \ref{phi3}) solved only in
case of presence of two scalar potentials  $V(\phi)$, and
$\omega(\phi)$. Here we have claimed that in the presence of
curvature term $\frac{k}{a^2}$, Eqs.(\ref{v2}, \ref{phi3}) may be
solved with  potential  $V(\phi)$. Hence, the following solution are
obtained \be \label{sol} \phi=t, \hspace{1cm} H=f(t) \ee
\\ From Eq.(\ref{phi3}) we get \be
\label{keq}\frac{k}{a^{2}}=3f^{2}(\phi)(\Omega_{\rm
\Lambda}-1)-2f'(\phi)+\frac{1}{M_{p}^{2}} \ee Substituting the above
$\frac{k}{a^2}$ into Eq.(\ref{v2}), we obtain the scalar potential
as \be \label{pottac111}V(\phi)=3M_{p}^{2}\Omega_{\Lambda}
f^2(\phi)+\frac{1}{2} \ee
  One can
check that the solution (\ref{sol}) satisfies the following scalar
field equation \be \label{phieq}-\ddot{\phi}-3H\dot{\phi}+V'(\phi)=0
\ee Therefore by the above condition, $f(\phi)$ in our model must
satisfy following relation \be \label{coneq} 3f(\phi)=V'(\phi)\ee
 In the other hand, using
Eqs.(\ref{holoda}, \ref{stateq}) we have \be
\label{v4}V(\phi)=\frac{3H^2 \Omega_{\rm \Lambda}}{16\pi
G}(\frac{4}{3}+\frac{2\sqrt{\Omega_{\rm
\Lambda}-c^2\Omega_{k}}}{3c}+\frac{b^2(1+\Omega_k)}{\Omega_{\rm
\Lambda}}) \ee \be \label{phi1} \dot{\phi}=\frac{H\sqrt{\Omega_{\rm
\Lambda}}}{2\sqrt{\pi G}}[-1+\frac{\sqrt{\Omega_{\rm
\Lambda}-c^2\Omega_{k}}}{c}+\frac{3b^2(1+\Omega_k)}{2\Omega_{\rm
\Lambda}}]^{1/2} \ee Using Eq.(\ref{phi1}), we can rewrite
Eq.(\ref{v4}) as \be \label{v5}V(\phi)=3 M_{p}^{2}\Omega_{\rm
\Lambda} H^2(1+\frac{\dot{\phi}^{2}}{6M_{p}^{2} H^2 \Omega_{\rm
\Lambda}}),\ee or in another form as following \be
\label{v5}V(\phi)=3 M_{p}^{2}\Omega_{\rm \Lambda}
[f^{2}(\phi)+\frac{1}{6M_{p}^{2}
 \Omega_{\rm \Lambda}}]=3M_{p}^{2}\Omega_{\Lambda}
f^2(\phi)+\frac{1}{2}\ee which is exactly the result
(\ref{pottac111}).\\
 From Eq.(\ref{phi3}) for the flat case we have
\be \label{flat}
2f'(\phi)=3f^{2}(\phi)(\Omega_{\Lambda}-1)+\frac{1}{M_{p}^{2}}\ee By
derivative of the above equation respect to $\phi$ we obtain \be
\label{flat1}
2f''(\phi)=6ff'(\phi)(\Omega_{\Lambda}-1)+3f^2\Omega_{\Lambda}'\ee
then \be \label{flat2} \Omega_{\Lambda}'=\frac{2f''}{3f}+\frac{2f'}
{f}(1-\Omega_{\Lambda})\ee Now using Eqs. (\ref{coneq}), (\ref{v5})
we have \be \label{flat3}
2f'\Omega_{\Lambda}+f\Omega_{\Lambda}'=\frac{1}{M_{p}^{2}}\ee
Substituting $\Omega_{\Lambda}'$ from Eq.(\ref{flat2}) into the
above equation we obtain \be \label{flat4}
2f''+6f'f-\frac{3f}{M_{p}^{2}}=0\ee Therefore, $f(\phi)$ must
satisfy the above equation in flat case. Using Maple software one
can obtain following relation \be \label{w}\int^{f(\phi)}\frac{2
dx}{W(c_1 e^{-(3x^{2}+1)})+1}=\phi+c_2 \ee where $W$ is the Lambert
$W$-function.\footnote{Consideration of Lambert $W$ function can be
traced back to J. Lambert around 1758, and later, it was considered
by L. Euler but it was recently established as a special function of
mathematics on its own\cite{31}.\\
The Lambert $W$ function is defined to be the function satisfying
\be \label{w2}W[z]e^{W[z]}=z \ee It is a multivalued function
defined in general for z complex and assuming values $W[z]$ complex.
If $z$ is real and $z < -1/e$, then $W[z]$ is multivalued complex.
If $z$ is real and $-1/e \leq z \leq 0$, there are two possible real
values of $W[z]$. The one real value of $W[z]$ is the branch
satisfying $\leq-1  W[z]$, denoted by $W_{0}[z]$, and it is called
the principal branch of the W function. The other branch is $W[z]
\leq -1$ and is denoted by $W_{-1}[z]$. If $z$ is real and $z \geq
0$, there is a single real value for $W[z]$ which also belongs to
the principal branch $W_{0}[z]$. Special values of the principal
branch of the Lambert $W$ function are $W_{0}[0] = 0 $and
$W_{0}[-1/e] = -1$. The Taylor series of $W_{0}[z]$ about $z = 0$
can be found using the Lagrange inversion theorem and is given by
\cite{32} \be \label{w2}W[z] = \sum_{1}^{\infty}=
\frac{(-1)^{n-1}n^{n-2}}{(n-1)!}z^{n} = z - z^{2} + \frac{3}{2}
z^{3} - \frac{8}{3} z^{4} + \frac{125}{24} z^{5} -\frac{54}{5} z^{6}
+ . . . . \ee The ratio test establishes that this series converges
if $|z| < 1/e$.}

\section{Squared speed for generalized Chaplygin gas and interacting holographic dark energy}
Here we introduce the squared speed of generalized Chaplygin gas as
\be \label{sp} v_{g}^{2}=\frac{dP_{\Lambda}}{d\rho_{\Lambda}} \ee
Using Eq.(14), we have \be \label{sp1}
v_{g}^{2}=\frac{A\alpha}{\rho^{\alpha+1}} \ee For $\alpha<0$, $A> 0$
or $\alpha >0$, $A <0$ (see recent paper \cite{lo}) ; $v_{g}^{2}<0$
, in this cases the generalized Chaplygin gas model is instable (see
also \cite{my}). The squared speed of interacting holographic dark
energy fluid is as \be \label{sp2}
v_{\Lambda}^{2}=\frac{dP_{\Lambda}}{d\rho_{\Lambda}}=
\frac{\dot{P}_{\Lambda}}{\dot{\rho}_{\Lambda}}\ee where \be
\label{sp3}\dot{P}_{\Lambda}= \dot{w}_{\rm
\Lambda}^{eff}\rho_{\Lambda}+w_{\rm
\Lambda}^{eff}\dot{\rho}_{\Lambda} \ee with \be \label{sp4}
\dot{w}_{\rm \Lambda}^{eff}= H \frac{dw_{\rm \Lambda}^{eff}}{dx} \ee
where $x=Ln a$. Using Eq.(27) and following equation \be
\label{evol}
\frac{d\Omega_{\Lambda}}{dx}=\frac{\dot{\Omega_{\Lambda}}}{H}=3\Omega_{\Lambda}(1+\Omega_{k}-\Omega_{\Lambda})
[\frac{1}{3}+\frac{2\sqrt{\Omega_{\rm
\Lambda}}}{3c}\frac{1}{\sqrt{|k|}}\rm cosn(\sqrt{|k|}\,R_{h}/a)] \ee
We can write \be \label{evol1} \dot{w}_{\rm
\Lambda}^{eff}=\frac{-H}{c\sqrt{\Omega_{\Lambda}- c^{2}\Omega_{k}}}
\Omega_{\Lambda}(1+\Omega_{k}-\Omega_{\Lambda})
[\frac{1}{3}+\frac{2\sqrt{\Omega_{\rm
\Lambda}}}{3c}\frac{1}{\sqrt{|k|}}\rm cosn(\sqrt{|k|}\,R_{h}/a)] \ee
Where we have assumed $\frac{d\Omega_{k}}{dx}=0$. Substituting the
above equation and Eq.(38) into Eq.(65) we obtain (to see the
non-interacting case relation refer to \cite{my} ) \be \label{sp5}
v_{\Lambda}^{2}=w_{\rm \Lambda}^{eff}-\frac{\dot{w}_{\rm
\Lambda}^{eff}}{3H(1+ w) } \ee From Eq.(69) one can see that ,
$\dot{w}_{\rm \Lambda}^{eff}<0$, also as we have mentioned in
section 3 if we take $c = 0.84$ , and taking $ \Omega_{\Lambda}=
0.73$, $ \Omega_{k} = 0.01$ for the present time, we obtain $w_{\rm
\Lambda}^{eff}=-1.007 $. One can see from Eq.(40)that in the phantom
phase where  $\dot{H}>0$
 ,$w+1 <0$ , hence we obtain a negative value for
 squared speed of interacting holographic fluid. Due to this the  interacting holographic fluid similare to
generalized Chaplygin gas is instable. In a recent paper Myung \cite{my} has shown that the perfect fluid for
holographic dark energy is classically unstable, our result show interacting  fluid of holographic
dark energy is also unstable. However, in contrast to the Chaplygin gas fluid where  the squared
speed is always non-negative, for the generalized  Chaplygin gas may be one can obtain negative value for
the squared speed. Hence the holographic interpretation for generalized
Chaplygin gas in contrast with Chaplygin gas   is not  problematic.

\section{Conclusions}
In order to solve cosmological problems and because the lack of our
knowledge, for instance to determine what could be the best
candidate for DE to explain the accelerated expansion of universe,
the cosmologists try to approach to best results as precise as they
can by considering all the possibilities they have. Within the
different candidates to play the role of the dark energy, the
Chaplygin gas, has emerged as a possible unification of dark matter
and dark energy, since its cosmological evolution is similar to an
initial dust like matter and a cosmological constant for late times.
Inspired by the fact that the Chaplygin gas possesses a negative
pressure, people \cite{mas} have undertaken the simple task of
studying a FRW cosmology of a universe filled with this type of
fluid.\\
In this paper, by considering an interaction between generalized
Chaplygin gas energy density and CDM, we have obtained the equation
of state for the interacting  generalized Chaplygin gas energy
density in the non-flat universe. Then we have considered a
correspondence between the holographic dark energy density and
interacting generalized  Chaplygin gas energy density in FRW
universe. Then we have reconstructed the potential of the
 scalar field which describe the generalized  Chaplygin cosmology.
Also we calculated the squared speeds of sound for generalized
Chaplygin gas and interacting holographic fluid, then we have shown
that interacting holographic fluid similar to the generalized
Chaplygin gas is instable.


\begin{thebibliography}{99}

\bibitem{exp} S. Perlmutter et al, Nature (London), 391, 51, (1998);
Knop. R et al., Astroph. J., 598, 102 (2003); A. G. Riess et al.,
Astrophy. J., 607, 665(2004); H. Jassal, J. Bagla and T.
Padmanabhan, Phys. Rev. D, 72, 103503 (2005).
\bibitem{cosmo} For review on cosmological constant
problem:  P. J. E. Peebles, B. Ratra,  Rev. Mod. Phys., 75, 559-606,
(2003);  J. Kratochvil, A. Linde, E. V. Linder, M. Shmakova,  JCAP,
0407, 001, (2004).
\bibitem{quint} R. R. Caldwell, R. Dave and P. J. Steinhardt, Phys. Rev. Lett., 80, (1998)
1582; I. Zlater, L. Wang and P. J. Steinhardt, Phys. Rev. Lett., 82,
(1999), 896; T. Chiba, gr-qc/9903094; M. S. Turner and M. White
Phys. Rev. D, 56, (1997), 4439.
\bibitem{phant} R. R. Caldwell, Phys.
Lett. B 545, 23, (2002); R. R. Caldwell, M. Kamionkowsky and N. N.
Weinberg, Phys, Rev, Lett, 91, 071301, (2003); S. Nojiri and S. D.
Odintsov, Phys. Lett., B {\bf562}, (2003), 147;
 S. Nojiri and S. D. Odintsov, Phys. Lett., B {\bf565}, (2003), 1;  S. Nojiri and S. D. Odintsov,
 Phys. Rev., D, 72, 023003, (2005); S. Nojiri, S. D. Odintsov, O. G. Gorbunova, J. Phys., A, 39, 6627,
 (2006); M. R. Setare, Phys. Lett. B {\bf641}, 130, (2006).
\bibitem{hologram}G. 't Hooft, gr-qc/9310026 ; L. Susskind, J. Math. Phys, {\bf36}, (1995),
6377.
\bibitem{myung}  Y. S. Myung, Phys. Lett. B {\bf610}, (2005), 18-22.
\bibitem{cohen} A. Cohen, D. Kaplan and A. Nelson, Phys. Rev. Lett
82, (1999), 4971.
\bibitem{fischler} W. Fischler and L. Susskind, hep-th/9806039.
\bibitem{miao} M. Li, Phy. Lett. B, 603, 1, (2004).
\bibitem{HDE} D. N. Vollic, hep-th/0306149; E. Elizalde, S. Nojiri,
S. D. Odintsov, and P. Wang,  Phys. Rev. D{\bf71}, 103504, (2005);
B. Guberina, R. Horvat, and H. Stefancic,  JCAP, 0505, 001, (2005);
B. Guberina, R. Horvat, and H. Nikolic, Phys. Lett. B{\bf 636}, 80,
(2006) ;H. Li, Z. K. Guo and Y. Z. Zhang, astro-ph/0602521; J. P. B.
Almeida and J. G. Pereira, gr-qc/0602103; D. Pavon and W. Zimdahl,
hep-th/0511053; Y. Gong, Phys. Rev., D, 70, (2004), 064029; B. Wang,
E. Abdalla, R. K. Su, Phys. Lett., B, 611, (2005);  M.~R.~Setare,
Phys. Lett. B {\bf644}, 99, (2007); M.~R.~Setare, 01, 023, JCAP
(2007); .~R.~Setare, and E. Vagenas, arXiv:0704.2070 [hep-th]
;M.~R.~Setare, arXiv:0705.3517 [hep-th]; K. H. Kim, H. W. Lee and Y.
S. Myung, Phys. Lett. {\bf B} 648, 107, (2007); K. H. Kim, H. W. Lee
and Y. S. Myung, arXiv:0706.2444 [gr-qc].
\bibitem{wmap}C. L. Bennett
et al., Astrophys. J. Suppl. 148, 1 (2003); D. N. Spergel,
Astrophys. J. Suppl. 148, 175, (2003).
\bibitem{ws}M. Tegmark et al., astro-ph/0310723.
\bibitem{miao2} Q. G. Huang and M. Li,  JCAP, 0408, 013, (2004).
\bibitem{guberina} B. Guberina, R. Horvat and H. Nikoli\'{c},
Phys. Rev. D {\bf72}, 125011, (2005).
\bibitem{intde}
  L.~Amendola,
    Phys.\ Rev.\ D {\bf 62}, 043511 (2000)
  [astro-ph/9908023];\\
    D.~Comelli, M.~Pietroni and A.~Riotto,
   Phys.\ Lett.\ B {\bf 571}, 115 (2003)
  [hep-ph/0302080];\\
    M.~R.~Setare, Phys. Lett. {\bf B642}, 421, (2006);\\
 M.~R.~Setare,  Eur. Phys. J. {\bf C50}, 991, (2007).
\bibitem{kmp} A. Yu. Kamenshchik, U. Moschella, and  V. Pasquier, Phys. Lett. {\bf B511}, 265, (2001).
\bibitem{jac} D. Bazeia, R. Jackiw, Ann. Phys. 270 (1998) 246; D. Bazeia,
Phys. Rev. D 59 (1999) 085007; R. Jackiw, A.P. Polychronakos,
Commun. Math. Phys. 207 (1999) 107; N. Ogawa, Phys. Rev. {\bf D62},
085023, (2000).
\bibitem{a}N. Bilic, G.B. Tupper and R.D. Viollier, Phys. Lett.
B535 (2002) 17; N. Bilic, G.B. Tupper and R.D. Vio- llier,
astro-ph/0207423.
\bibitem{b}M. Bordemann and J. Hoppe, Phys. Lett. B317 (1993)
315; J.C. Fabris, S.V.B. Gonsalves and P.E. de Souza, Gen. Rel.
Grav. 34 (2002) 53.
\bibitem{c}M.C. Bento, O. Bertolami and A.A. Sen, Phys. Lett.
B575 (2003) 172.
\bibitem{Kim:2005at} H. Kim, H. W. Lee and Y. S. Myung,
Phys. Lett. B {\bf 632}, 605, (2006).

\bibitem{WGA} B. Wang, Y. Gong, and E. Abdalla, Phys. Lett. B {\bf 624}, 141
(2005).
\bibitem{set1}M. R. Setare, Phys. Lett. {\bf B642}, 1, (2006).
\bibitem{set2}M. R. Setare, Jingfei Zhang, Xin Zhang, JCAP {\bf 0703}, 007, (2007).
\bibitem{set4}M. R. Setare, Phys. Lett. {\bf B648}, 329, (2007).
\bibitem{shap}I.L. Shapiro, J. Sola,  C. Espana-Bonet, and P.
Ruiz-Lapuente,  Phys. Lett. {\bf B574}, 149, (2003).
\bibitem{obsnonflat}
  Y.~G.~Gong, B.~Wang and Y.~Z.~Zhang,
  Phys.\ Rev.\ D {\bf 72}, 043510 (2005).
    \bibitem{odi1}S. Nojiri,  and S. D. Odintsov, Gen. Rel. Grav. {\bf 38}, 1285,
(2006).
\bibitem{31}R. M. Corless, G. H. Gonnet, D. E. G. Hare, D. J. Jeffrey and D. E. Knuth, Adv.
Comput. Math. 5, 329 (1996).
\bibitem{32}R. M. Corless, D. J. Jeffrey and D. E. Knuth, Proceedings ISSAC '97, 197 (1997).
\bibitem{odi2}S. Capozziello, S. Nojiri,  and S. D. Odintsov,  Phys. Lett. {\bf B632}, 597, (2006);
S. Nojiri,  and S. D. Odintsov, hep-th/0611071; S. Nojiri, S. D.
Odintsov, and H. Stefancic, Phys. Rev. {\bf D74}, 086009, (2006).
\bibitem{mas}V. Gorini, A. Kamenshchik, U. Moschella
and V. Pasquier, gr-qc/0403062.
\bibitem{lo}M. Bouhmadi-L´opez,  P. F. Gonz´alez-Diaz,  and P. Martin-Moruno , 0707.2390[gr-qc].
\bibitem{my}Y. S. Myung,  Phys. Lett. {\bf B} 652, 223, (2007).

\end{thebibliography}
\end{document}